\documentclass{article}

\usepackage{arxiv}

\usepackage[utf8]{inputenc} 
\usepackage[T1]{fontenc}    
\usepackage{hyperref}       
\usepackage{url}            
\usepackage{booktabs}       
\usepackage{amsfonts}       
\usepackage{nicefrac}       
\usepackage{microtype}      
\usepackage{lipsum}
\usepackage{graphicx}
\usepackage{cite}

\usepackage{multirow}
\usepackage{subcaption}
\graphicspath{ {./images/} }

\title{Differences in Perspective on Inertial Measurement Unit Sensor Integration in Myoelectric Control}

\author{
 Evan Campbell \\
  Institute of Biomedical Engineering\\
  University of New Brunswick\\
  \texttt{evan.campbell1@unb.ca} \\
   \And
 Angkoon Phinyomark \\
  Institute of Biomedical Engineering\\
  University of New Brunswick\\
  \texttt{aphinyom@unb.ca} \\
  \And
  Erik Scheme \\
  Institute of Biomedical Engineering\\
  University of New Brunswick\\
  \texttt{escheme@unb.ca} \\ \\
}

\begin{document}
\maketitle
\begin{abstract}

Recent human computer-interaction (HCI) studies using electromyography (EMG) and inertial measurement units (IMUs) for upper-limb gesture recognition have claimed that inertial measurements alone result in higher classification accuracy than EMG. 
In biomedical research such as in prosthesis control, however, EMG remains the gold standard for providing gesture specific information, exceeding the performance of IMUs alone.
This study, therefore, presents a preliminary investigation of these conflicting claims between these converging research fields.
Previous claims from both fields were verified within this study using publicly available datasets.
The conflicting claims were found to stem from differences in terminology and experimental design. Specifically, HCI studies were found to exploit positional variation to increase separation between similar hand gestures. Conversely, in clinical applications such as prosthetics, position invariant gestures are preferred.
This work therefore suggests that future studies explicitly outline experimental approaches to better differentiate between gesture recognition approaches.
\end{abstract}


\section{Introduction}
\quad Gesture recognition using electromyography (EMG) pattern recognition has a long history of use in biomedical and clinical applications, such as myoelectric control of prosthetic devices and other assistive or rehabilitative technologies.
These devices leverage residual motor function to enhance quality of life limited by neurological (stroke~\cite{HoStroke}) or physical impairment (amputation~\cite{CampbellNER}).
The emerging interest in hand gesture recognition as a general human-computer interface (HCI) for consumer applications, such as virtual reality, has large commercial incentives and has therefore accelerated in recent years.
The use of wrist- or forearm-worn EMG devices combined with inertial sensors (i.e., accelerometer (ACC), magnetometer (MAG), or gyroscope (GYR)) have demonstrated the potential of such gesture recognition interfaces during offline classification studies~\cite{jiang2017feasibility}.
These multi-modal devices have been validated in both biomedical and general HCI studies; however, the conditions of gesture elicitation differ between the two applications.

Biomedical applications of EMG pattern recognition typically require accurate recognition of physiologically appropriate gestures that are robust to variability of daily-living; simply put, the gestures should be reliably decoded regardless of limb posture and contraction intensity, among other factors \cite{scheme2011electromyogram}.
Limb posture and contraction intensity variability degrades the usability of clinical EMG pattern recognition systems meaningfully, as gesture recognition accuracies were found to decrease on the order of 13\% and 20\% for these factors, respectively, across several studies \cite{CampbellSensors}.
Interventions in the form of training strategies \cite{Yang2017}, algorithmic solutions \cite{Khushaba2014NN}, or multi-sensor approaches \cite{fougner2011multi} have lessened this degradation and led to more reliable use of myoelectric control.
Multi-sensor approaches using EMG and ACC measurements from many positions have altogether removed degradation caused by static limb positions in recorded positions by sequential use of a position-classifier using ACC, followed by a position-specific EMG classifier for gesture recognition~\cite{fougner2011multi}.
No application other than position recognition, however, has been validated for non-mechanomyographic ACC measurements within clinical EMG pattern recognition studies.

Alternatively, general HCI applications of EMG pattern recognition desire accurate recognition of distinct gestures; the gestures in these application are no longer required to be invariant to daily-living variability and may selectively harness position variability to become more distinct.
Consequently, inertial sensors have been found to outperform EMG sensors in terms of gesture recognition accuracy~\cite{jiang2017feasibility, krasoulis2017improved, khushaba2018spatio,koch2019inhomogeneously}.
For instance, gesture recognition using MAG achieved 93\% accuracy across 40 motion classes, whereas EMG achieved only 65\%.
The different interpretation of the application and value of inertial measurements between biomedical and HCI studies is a current area of confusion in the field that warrants further clarification.

This paper aims to highlight the main differences between biomedical and HCI studies of EMG pattern recognition by examining the differences between gesture elicitation studies.
Specifically, this study focused on the differences in the gestures performed and the differences in the use of inertial information.
Differences in gestures are presented through visualization of signals, whereas the differing use of inertial information is presented through classification outcomes using EMG and ACC feature sets.

\section{METHODS}
\label{sec:methods}

\begin{figure}[t]
    \centering
    \begin{tabular}[b]{| c c | c c c |}
    \toprule
    \multicolumn{2}{|c|}{Biomedical dataset} & \multicolumn{3}{c|}{HCI dataset} \\ \midrule
        \begin{subfigure}[b]{0.18\textwidth}
            \centering
            \smallskip
            \includegraphics[width=\linewidth]{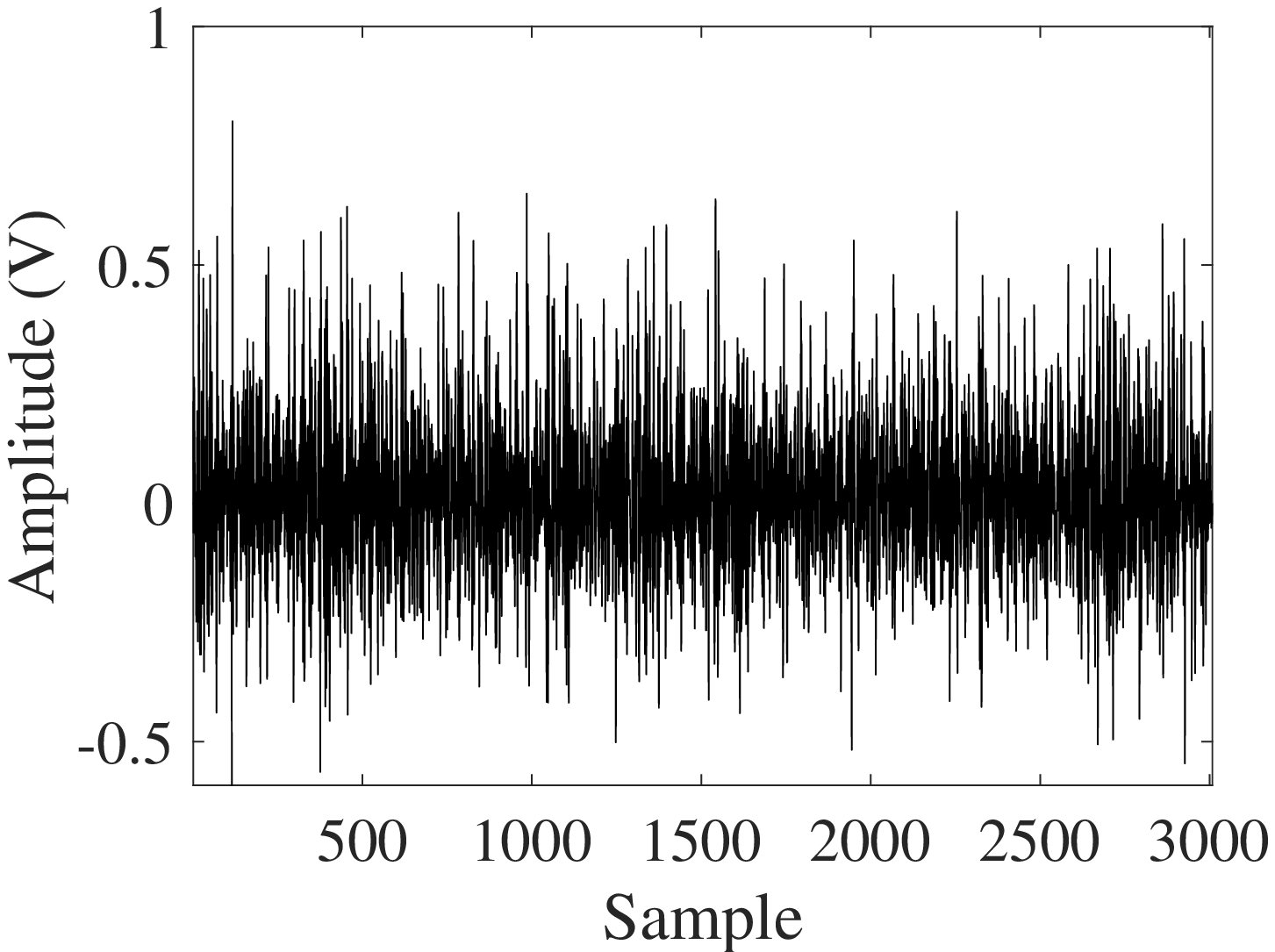}
            \caption{WF, EMG} 
        \end{subfigure}
        &
        \begin{subfigure}[b]{0.18\textwidth}
            \centering
            \smallskip
            \includegraphics[width=\linewidth]{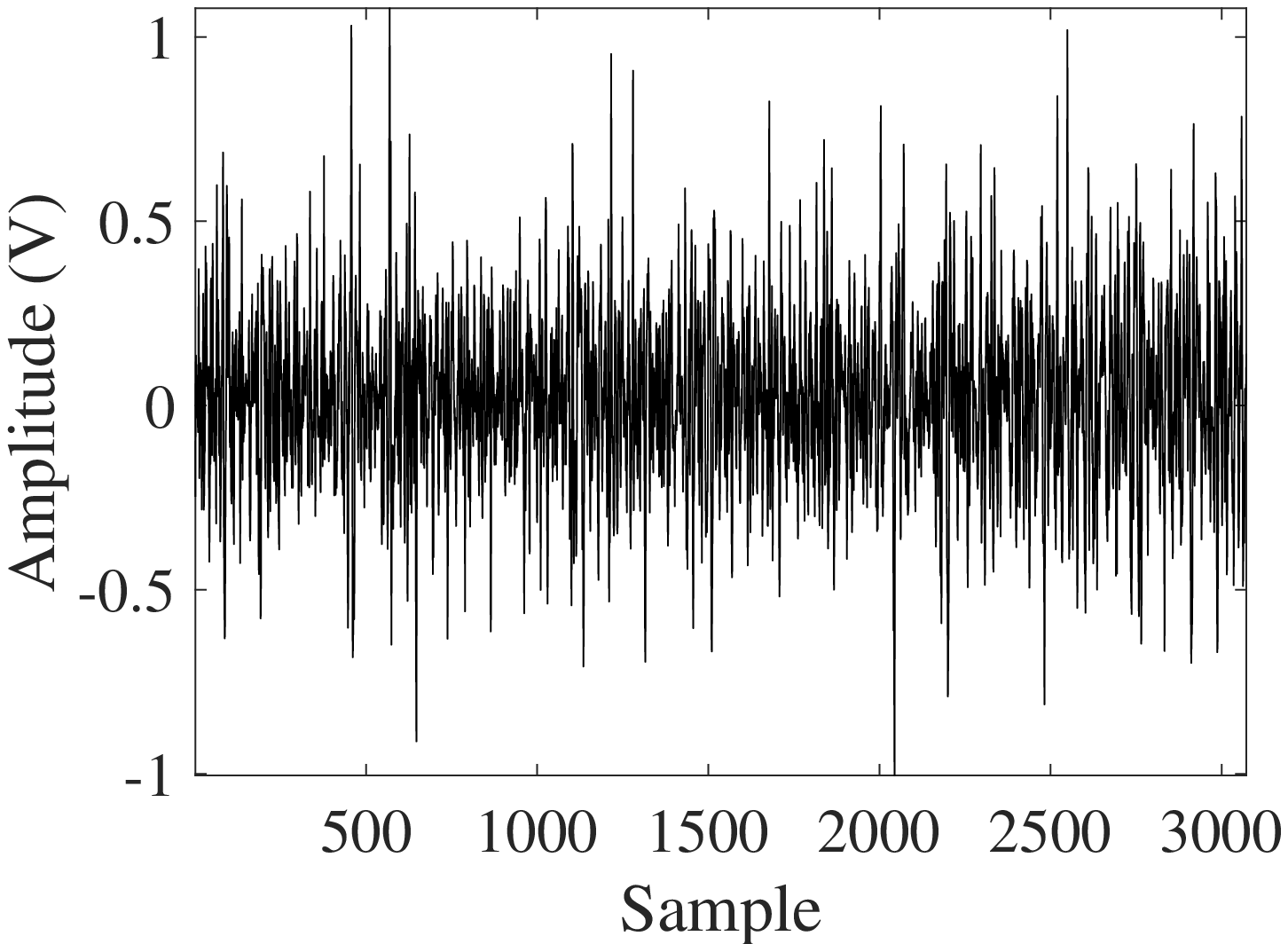}
            \caption{WP, EMG} 
        \end{subfigure}
        &
        \begin{subfigure}[b]{0.18\textwidth}
            \centering
            \smallskip
            \includegraphics[width=\linewidth]{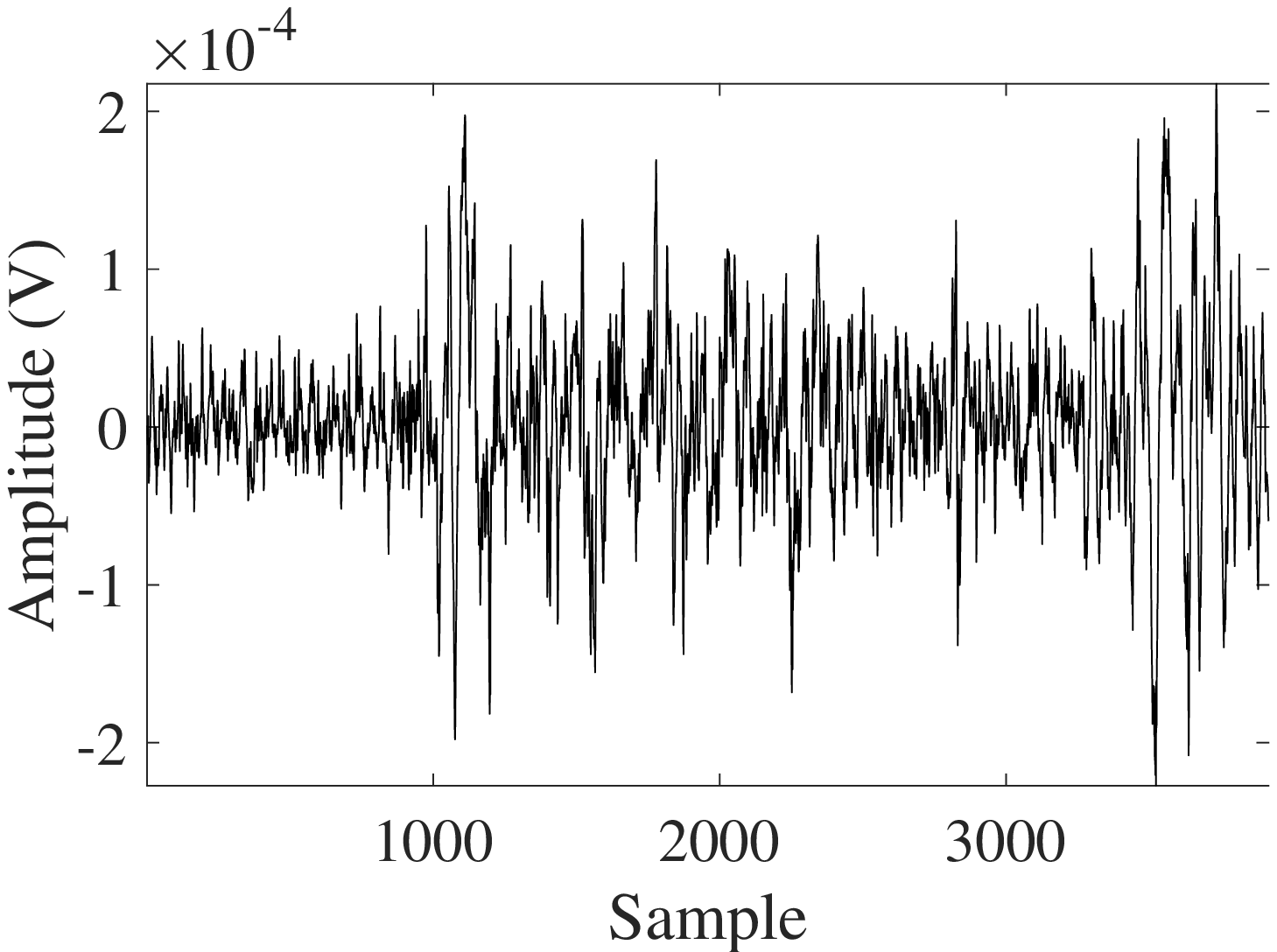}
            \caption{WF, EMG} 
        \end{subfigure}
        &
        \begin{subfigure}[b]{0.18\textwidth}
            \centering
            \smallskip
            \includegraphics[width=\linewidth]{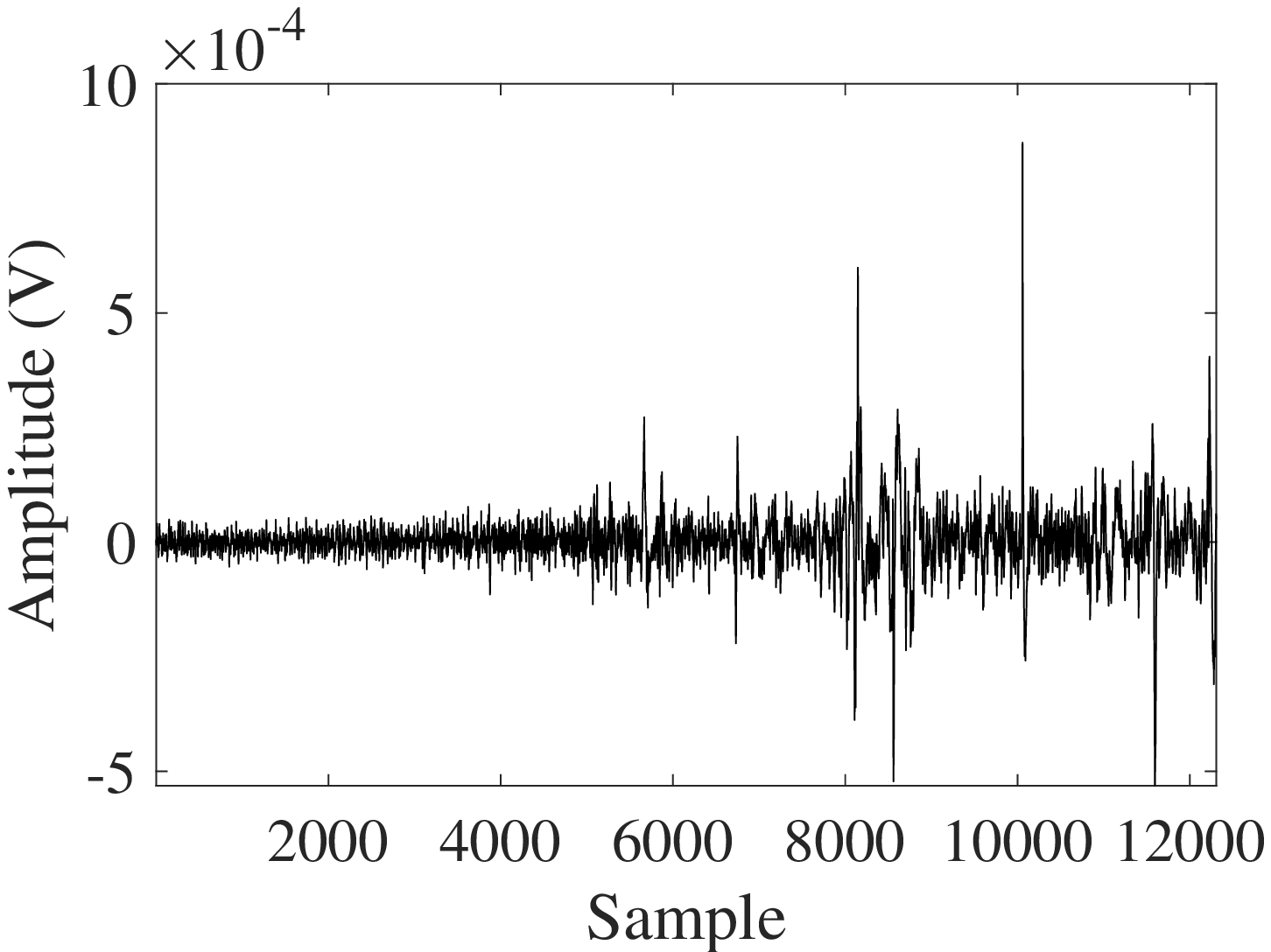}
            \caption{WP, EMG} 
        \end{subfigure}
        &
        \begin{subfigure}[b]{0.18\textwidth}
            \centering
            \smallskip
            \includegraphics[width=\linewidth]{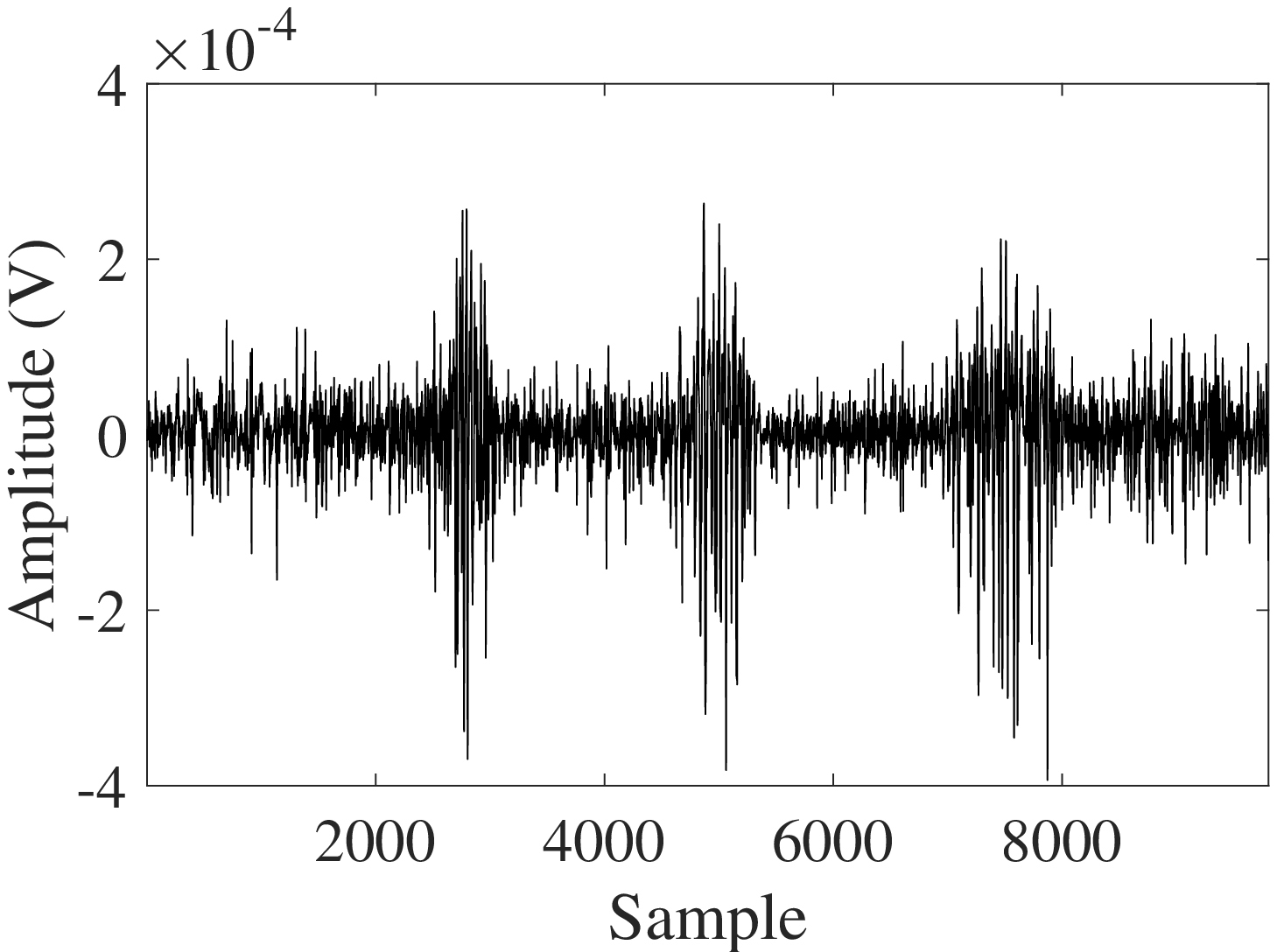}
            \caption{TS, EMG} 
        \end{subfigure} \\
        
       \begin{subfigure}[b]{0.18\textwidth}
            \centering
            \smallskip
            \includegraphics[width=\linewidth]{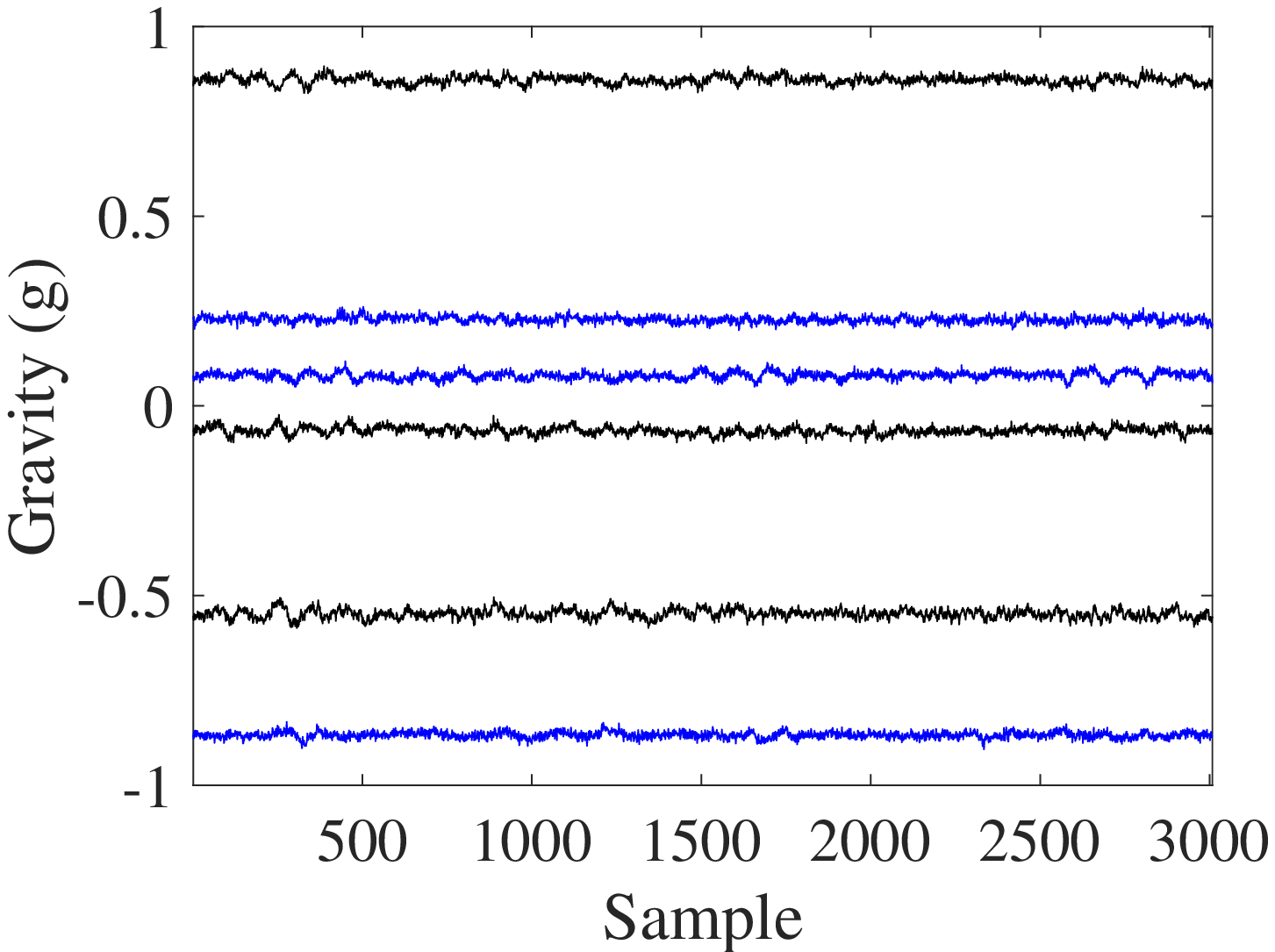}
            \caption{WF, ACC} 
        \end{subfigure}
        &
        \begin{subfigure}[b]{0.18\textwidth}
            \centering
            \smallskip
            \includegraphics[width=\linewidth]{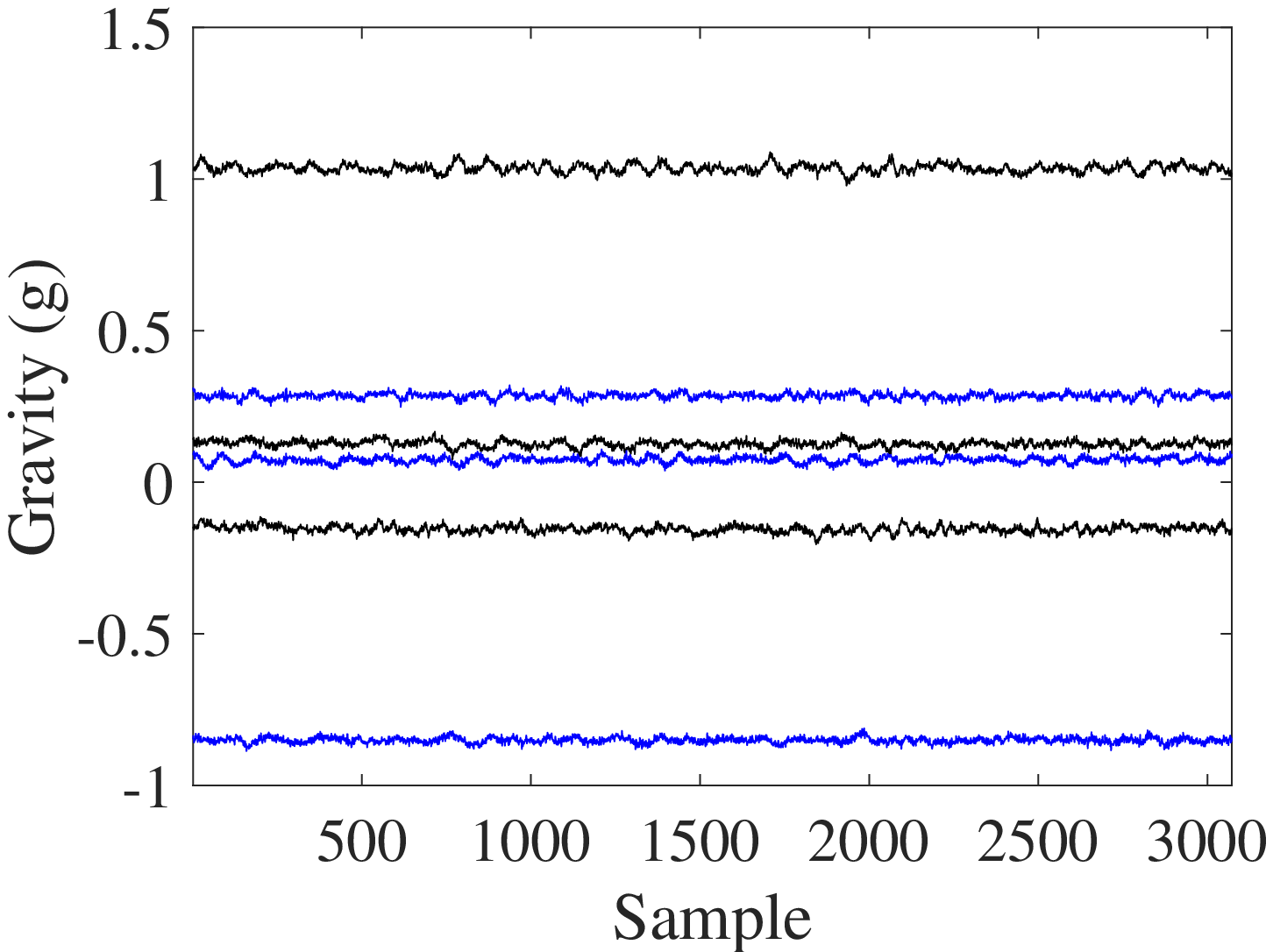}
            \caption{WP, ACC} 
        \end{subfigure}
        &
        \begin{subfigure}[b]{0.18\textwidth}
            \centering
            \smallskip
            \includegraphics[width=\linewidth]{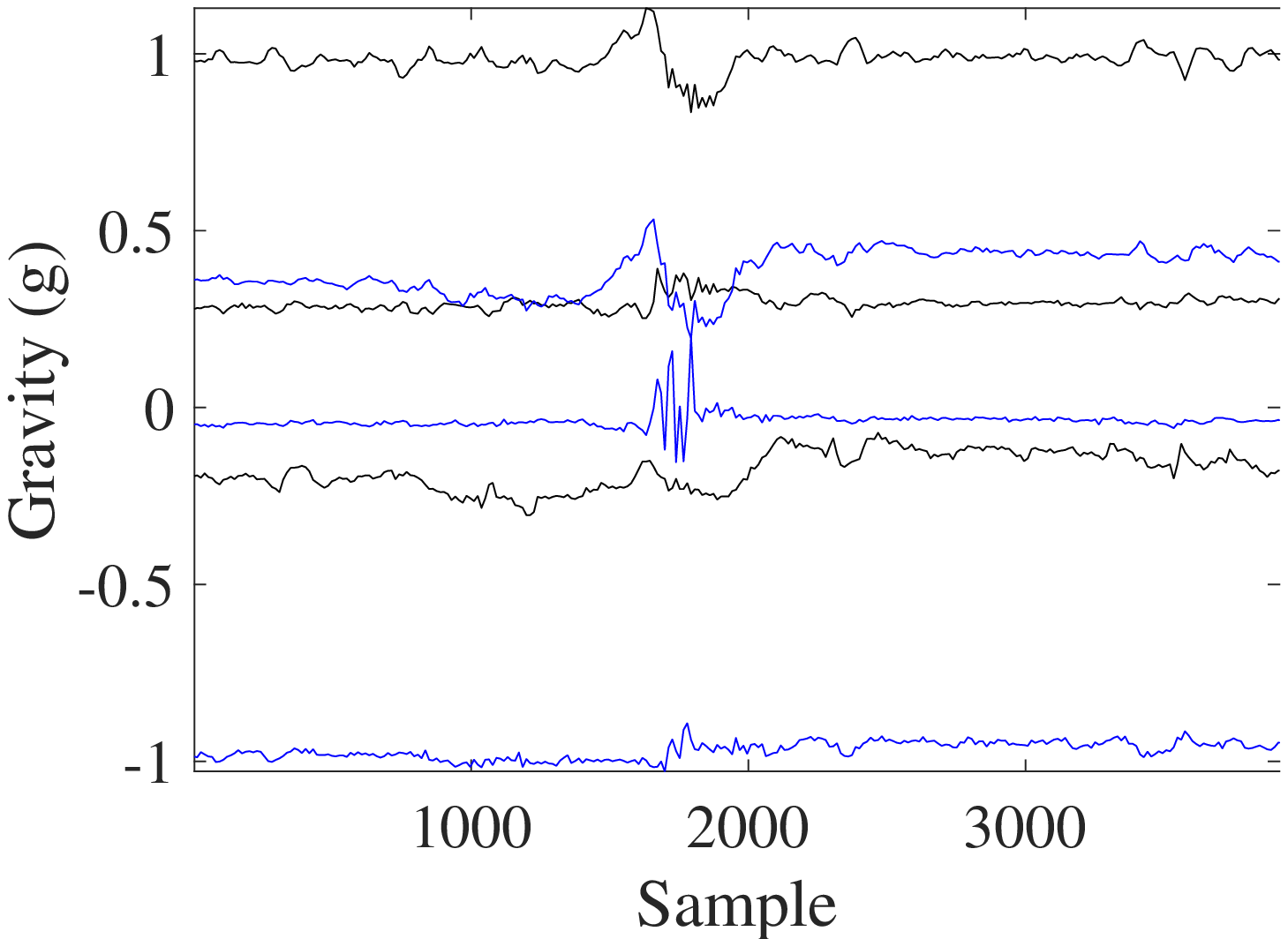}
            \caption{WF, ACC} 
        \end{subfigure}
        &
        \begin{subfigure}[b]{0.18\textwidth}
            \centering
            \smallskip
            \includegraphics[width=\linewidth]{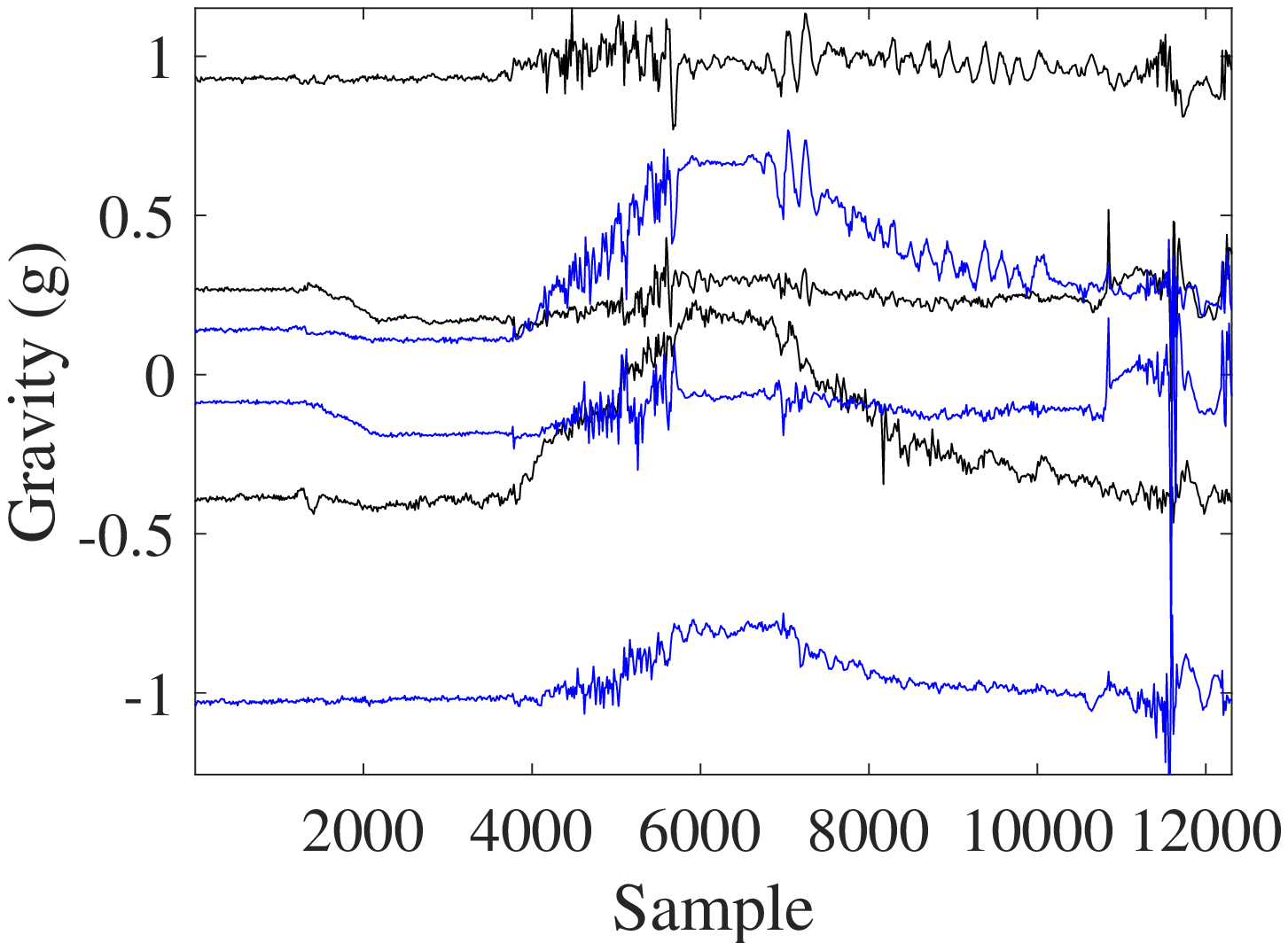}
            \caption{WP, ACC} 
        \end{subfigure}
        &
        \begin{subfigure}[b]{0.18\textwidth}
            \centering
            \smallskip
            \includegraphics[width=\linewidth]{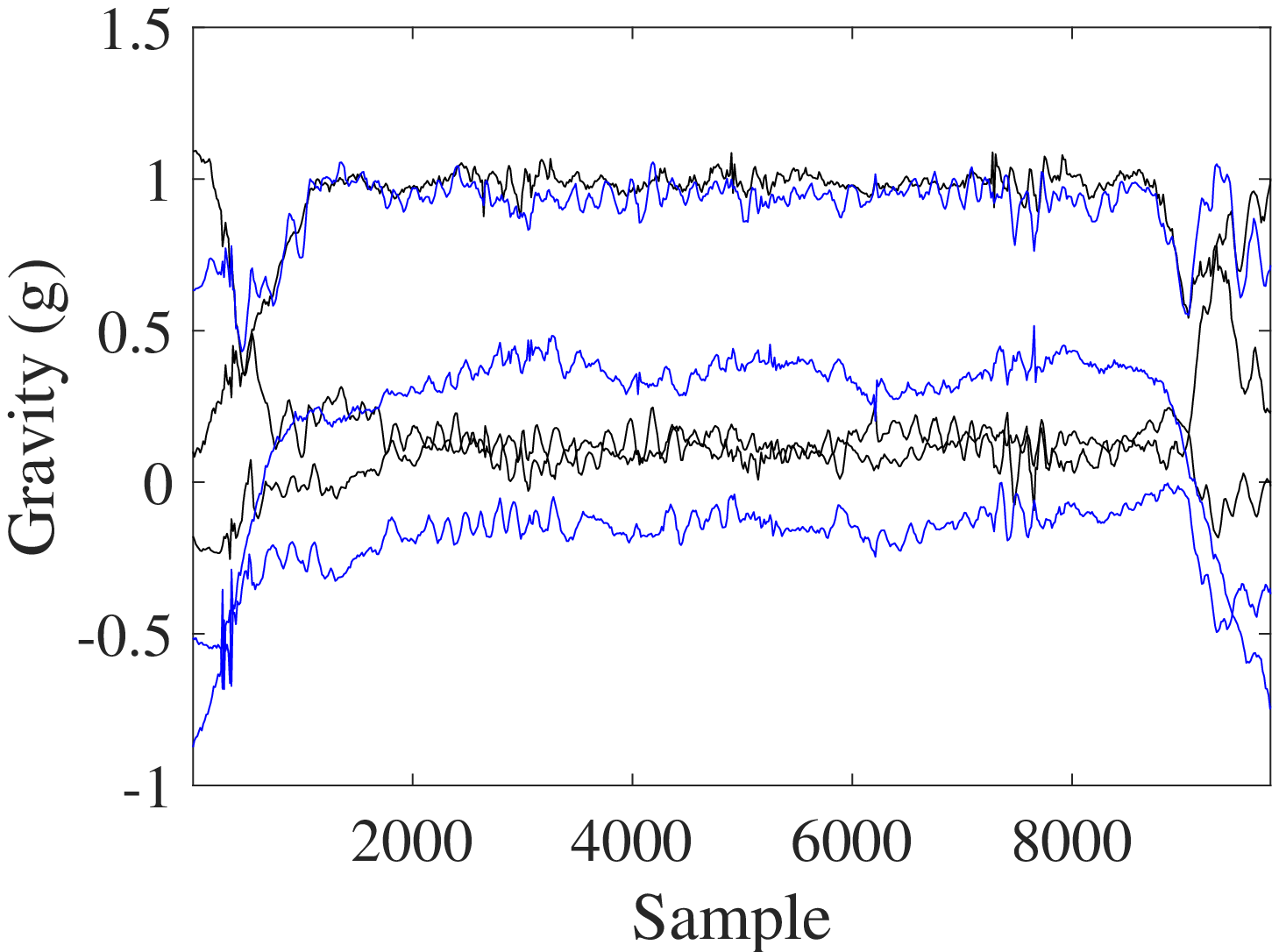}
            \caption{TS, ACC} 
        \end{subfigure} \\ \bottomrule
        
    \end{tabular}
    \caption{EMG and ACC measurements (unfiltered) from the biomedical and HCI datasets. The first row contains the EMG elicited during wrist flexion (WF), wrist pronation (WP), and a turning screw (TS) gesture. The second row contains the accelerometer readings for the same contractions, where the black lines represent the x, y, and z components of a forearm mounted sensors and the blue lines represent measurements simultaneously taken at the biceps.}
        \label{fig:signals}
\end{figure}

\subsection{Datasets}
Two public datasets were adopted to represent biomedical and HCI gesture recognition studies; the Fougner~\cite{fougner2011multi} and NinaPro7~\cite{krasoulis2017improved} datasets, respectively. 
The biomnedical dataset was collected using 8 bipolar Ag/AgCl electrodes (EMG) and 2 tri-axis accelerometers.
Twelve intact-limbed subjects performed 6 motions (wrist flexion, wrist extension, wrist pronation, wrist supination, hand close, and pinch grip) and no motion, where each motion was repeated 10 times in 5 different static limb positions.
The HCI dataset contained 12 bipolar Ag/AgCl electrodes and 12 tri-axis accelerometers.
Twenty intact-limbed subjects performed 40 dynamic gestures (8 finger gestures, 9 wrist gestures, and 23 grasping gestures), where each motion was repeated 6 times with limb position unspecified.
The gestures of the HCI dataset were segmented into 3 gesture sets: HCI-A, a set matching the biomedical dataset gestures, HCI-B, a subset containing 8 finger gestures, and HCI-C, a subset containing 23 grasp gestures.
A sample of EMG and ACC signals from both datasets is given in Fig. \ref{fig:signals}.

\subsection{Data preparation}

The EMG signals from both datasets were pre-processed by a 60 Hz or 50 Hz notch filter and 20-450 Hz bandpass filter to remove power-line interference and motion artefacts, respectively.
The ACC signals were pre-processed using 1 Hz low-pass filters, to remove accompanying sensor noise from measurements.
Both EMG and ACC signals of all channels were segmented into overlapping windows using window length and increment of 200 and 100 ms, respectively.

Features were extracted from each window to create 2 EMG and 2 ACC feature sets.
The EMG feature sets were the Hudgins' time-domain (TD) feature set \cite{Hudgins1993} (mean absolute value, zero crossings, slope sign change, and waveform length), and the time domain power spectral descriptors (TDPSD) feature set \cite{Khushaba2014NN}. 
The ACC feature sets were the median feature set (MED) and root mean square (RMS) feature set.

\subsection{Classification problems}

The four feature sets of all four datasets (biomedical, HCI-A, HCI-B, and HCI-C) were used in three classification tasks, where applicable, to validate claims proposed by previous studies.
\begin{enumerate}
    \item Multi-gesture position classification: Classifiers were trained  with feature vectors from all gestures with the class label selected as the position of the gesture. Only the biomedical dataset was used for this analysis, as the HCI dataset did not specify any specific limb positions.
    \item Within-position gesture classification: Classifiers were trained with feature vectors from an individual position with the class label being the associated gesture. This process was repeated for all positions in the case of the biomedical dataset and only a single position was assumed for the HCI dataset. 
    \item Sequential classification: Classifiers were first trained following the multi-gesture position classification task, where feature vectors were used to predict position. Subsequently, the position was used to select the appropriate position-specific gesture classifier, as was conducted in the within-position gesture classification task. As the HCI datasets did not provide labelled positions, they were excluded from this task.
\end{enumerate}

All classification tasks were performed using within-subject leave-one-trial-out cross-validation using linear discriminant analysis (LDA), quadratic discriminant analysis (QDA), \textit{k}-nearest neighbours (\textit{k}NN, \textit{k}=5), and random forest (RF, 10 trees) classifiers.
Accuracies are presented as mean + standard deviation, where the mean accuracy is the mean accuracy across all subjects and cross-validations, and the standard deviation is the standard deviation across subjects.

\section{RESULTS}
\label{sec:results}

\quad The multi-gesture position recognition results using ACC MED, ACC RMS, EMG TD, and EMG TDPSD feature sets are shown in Table \ref{table:position} for the biomedical dataset.
The within-position gesture recognition results of the biomedical and HCI datasets were presented in Table \ref{table:gesture}.
The LDA classifier was found to have the best performance among classifiers for all datasets in this latter classification task, again justifying its predominant use in myoelectric control \cite{campbell2019linear}.
The EMG TD feature set was found to be best for the biomedical dataset whereas ACC MED was found to best for all HCI datasets.
Further inspection of the performance of the EMG TD feature set with the LDA classifier is provided through the confusion matrices of the biomedical and HCI-A dataset in Table \ref{table:confusion}.
Conversely, Table \ref{table:confusion} shows a similar confusion matrix using the best feature set determined for the HCI dataset (ACC MED).
Finally, the results of sequential classification of gestures from multiple-positions are presented in Table \ref{table:sequential}.

\begin{table}[t]
\centering
 \caption{Multi-gesture position recognition accuracy (mean+std of subjects) across positions of the biomedical dataset}
\label{table:position}

    \begin{tabular}{l|cc|cc}
    \toprule
          \multirow{2}{*}{Classifier}&  \multicolumn{2}{c}{\underline{ACC}} & \multicolumn{2}{c}{\underline{EMG}} \\
                      & MED &  RMS  &  TD  & TDPSD \\ 
        \midrule
         LDA          &    99.9+0.3 &96.3+5.2 &63.0+9.7 &62.3+8.0 \\ 
         QDA          & 99.9+0.1 &98.4+1.8 &67.8+8.9 &66.0+7.6 \\ 
         \textit{k}NN & \textbf{100.0+0.0} &98.0+2.4 &66.8+8.1 &54.8+8.4 \\ 
         RF           & 99.5+0.6 &96.3+3.2 &66.8+8.4 &63.0+8.5 \\ 
    \bottomrule
    \end{tabular}
    
\end{table}

\begin{table}[b]
\centering
 \caption{Within-position gesture recognition rates across positions}
\label{table:gesture}

    \begin{tabular}{ll|cc|cc}
    \toprule
        \multirow{2}{*}{Dataset} &  \multirow{2}{*}{Classifier}&  \multicolumn{2}{c}{\underline{ACC}} & \multicolumn{2}{c}{\underline{EMG}} \\
        & & MED &  RMS  &  TD  & TDPSD \\ 
        \midrule
        \multirow{4}{*}{Bio}   &  LDA          &69.8+4.4 &65.8+4.5 &\textbf{96.2+0.7} &96.0+0.4 \\ 
                                    &  QDA           &66.4+4.8 &64.3+5.1 &95.1+0.8 &94.2+0.5 \\ 
                                    &  \textit{k}NN  &63.8+5.6 &60.8+5.1 &94.3+0.9 &85.8+1.2 \\ 
                                    &   RF           &61.2+4.9&59.2+3.3 &92.9+0.7 &91.6+0.9 \\ 
                                 \midrule
        \multirow{4}{*}{HCI-A}      &  LDA           &\textbf{97.1+1.5} &96.6+1.9 &89.1+3.5 &91.1+2.7 \\ 
                                    &  QDA           &93.8+3.8 &89.0+5.5 &82.9+5.3 &68.4+7.0 \\ 
                                    &  \textit{k}NN  &94.2+2.6 &94.6+2.4 &82.8+4.5 &70.1+4.8 \\ 
                                    &   RF           &92.0+3.8 &92.9+2.5 &85.4+3.6 &82.3+3.8 \\ 
                                    \midrule
        \multirow{4}{*}{HCI-B}      &  LDA           &\textbf{94.4+4.0} &94.2+4.1 &84.7+8.1 &87.5+8.6 \\ 
                                    &  QDA           &88.5+8.5 &84.4+8.5 &75.0+8.4 &53.1+10.4 \\ 
                                    &  \textit{k}NN  &87.7+8.8 &87.9+8.6 &68.3+9.1 &50.6+9.2 \\ 
                                    &   RF           &84.2+6.9 &84.4+7.1 &78.4+7.0 &73.0+8.4 \\
                                    \midrule
        \multirow{4}{*}{HCI-C}      &  LDA           &\textbf{89.1+4.4} &84.5+6.6 &66.5+8.5 &71.9+8.5 \\ 
                                    &  QDA           &87.9+8.1 &84.1+8.9 &60.9+9.6 &45.9+8.8 \\ 
                                    &  \textit{k}NN  &80.6+9.1 &81.7+9.2 &52.0+9.8 &34.1+7.1 \\ 
                                    &   RF           &77.9+8.9 &78.2+8.9 &62.3+8.6 &54.2+8.0 \\ 
    \bottomrule
    \end{tabular}

\end{table}

\begin{table}[h]
\centering
 \caption{Confusion matrix of the gesture classification accuracy (\%) of the biomedical and HCI-A datasets using EMG TD and ACC MED feature sets using the LDA classifier}
\label{table:confusion}

    \begin{tabular}{l|cccccc|cccccc}
    \toprule
        \multicolumn{13}{c}{\textbf{EMG TD}} \\ \midrule
          \multirow{2}{*}{True Label}& \multicolumn{6}{c|}{\underline{Predicted Label - Biomedical}} & \multicolumn{6}{c}{\underline{Predicted Label - HCI-A}}\\
                     & WF & WE& WP& WS& PO&PI& WF & WE& WP& WS& PO&PI\\
        \midrule
         WF &\textbf{97.4} &0.0 &1.0 &0.9 &0.5 &0.2 & \textbf{94.4} &0.6 &0.7 &1.0 &1.8 &1.5 \\
         WE &0.2 &\textbf{96.6} &0.1 &0.7 &1.8 &0.5 &0.2 &\textbf{86.3} &7.4 &1.7 &1.6 &2.8  \\
         WP &0.2 &0.3 &\textbf{96.2 }&1.9 &0.8 &0.5 &0.0 &10.2 &\textbf{83.5} &2.3 &1.0 &3.0 \\ 
         WS &0.1 &0.0 &1.0 &\textbf{97.8} &0.7 &0.4 &0.3 &1.0 &2.6 &\textbf{82.7} &10.9 &2.4 \\
         PO &0.1 &0.3 &0.4 &0.5 &\textbf{95.2} &3.5 &0.3 &1.1 &1.1 &10.8 &\textbf{85.0} &1.7 \\
         PI &0.3 &0.1 &0.5 &0.5 &4.0 &\textbf{94.6} &0.1 &0.6 &0.5 &0.4 &0.1 &\textbf{98.4}  \\

    \midrule
    \multicolumn{13}{c}{\textbf{ACC MED}} \\ \midrule
              \multirow{2}{*}{True Label}& \multicolumn{6}{c|}{\underline{Predicted Label - Biomedical}} & \multicolumn{6}{c}{\underline{Predicted Label - HCI-A}}\\
                     & WF & WE& WP& WS& PO&PI& WF & WE& WP& WS& PO&PI\\
    \midrule
         WF &\textbf{64.3} &15.8 &2.1 &3.2 &8.8 &5.7 & \textbf{99.9} &0.1 &0.0 &0.0 &0.0 &0.0 \\ 
         WE &15.8 &\textbf{59.6} &1.0 &3.3 &10.8 &9.6 & 1.0 &\textbf{97.5} &1.5 &0.0 &0.0 &0.0 \\ 
         WP &4.0 &2.8 &\textbf{87.2} &1.0 &3.9 &1.1 & 0.4 &6.4 &\textbf{92.7} &0.5 &0.0 &0.1 \\ 
         WS &4.4 &5.3 &0.6 &\textbf{83.9} &3.6 &2.3 & 0.0 &0.4 &0.6 &\textbf{95.1} &3.9 &0.0 \\ 
         PO &10.0 &9.8 &2.7 &2.7 &\textbf{50.6} &24.3 &0.0 &0.0 &0.0 &4.8 &\textbf{95.2} &0.0 \\
         PI &8.8 &11.6 &2.3 &3.5 &25.5 &\textbf{48.2} & 0.0 &0.0 &0.0 &0.0 &0.0 &\textbf{100.0}\\ \bottomrule
    \end{tabular}

\end{table}

\begin{table}[h]
\centering
 \caption{Sequential gesture recognition accuracy (mean+std of subjects) of the biomedical dataset}
\label{table:sequential}

    \begin{tabular}{ll|cccc}
    \toprule
        \multicolumn{2}{c}{\underline{Feature Set}} & \multicolumn{4}{c}{\underline{Classifier}} \\
        Position&  Gesture&  LDA & QDA &\textit{k}NN & RF \\
        \midrule
        \multirow{5}{*}{ACC MED}            & ACC MED   & 65.5+17.2 &62.4+15.8 &60.2+15.5 &58.2+15.1 \\ 
                                            & ACC RMS   &61.9+16.4 &60.5+15.7 &57.5+15.5 &55.8+15.4 \\ 
                                            &\textbf{EMG TD}    & \textbf{96.0+3.2} &94.7+4.3 &93.9+4.2 &92.5+4.0 \\ 
                                            & EMG TDPSD &95.6+3.5 &93.6+4.0 &84.7+5.5 &91.0+4.6 \\  \midrule
                                    
        \multirow{5}{*}{ACC RMS}            & ACC MED   & 63.8+15.9 &61.9+15.5 &59.5+14.8 &57.2+14.8 \\ 
                                            & ACC RMS   & 60.5+15.2 &60.0+15.3 &56.7+14.9 &54.1+14.3 \\ 
                                            & EMG TD    & 95.5+3.1 &94.4+4.2 &93.6+4.3 &91.5+4.2 \\ 
                                            & EMG TDPSD & 95.2+3.4 &93.3+3.8 &84.3+5.4 &90.6+4.5 \\  \midrule
                                    
        \multirow{5}{*}{EMG TD}             & ACC MED   & 50.2+12.0 &49.7+11.1 &51.1+13.0 &47.3+11.5 \\ 
                                            & ACC RMS   & 46.7+11.0 &47.4+11.0 &46.6+12.5 &44.1+11.6 \\ 
                                            & EMG TD    & 93.4+4.8 &92.8+5.3 &94.1+4.7 &91.2+4.7 \\ 
                                            & EMG TDPSD & 93.2+4.7 &92.2+4.5 &84.3+4.3 &89.8+5.0 \\  \midrule
                                    
        \multirow{5}{*}{EMG TDPSD}          & ACC MED   & 49.7+11.7 &49.1+11.2 &47.0+10.8 &46.5+11.5 \\ 
                                            & ACC RMS   & 45.8+10.5 &47.3+11.1 &42.0+9.9 &43.4+10.8 \\ 
                                            & EMG TD    & 93.3+5.0 &92.8+5.6 &91.2+5.4 &90.2+5.1 \\ 
                                            & EMG TDPSD & 93.5+4.7 &92.1+4.5 &83.1+4.3 &89.2+5.3 \\

    \bottomrule
    \end{tabular}

\end{table}

\section{DISCUSSION}
\label{sec:discussion}

\quad This study corroborates the use of ACC as an accompanying modality in biomedical/clinical applications to achieve positional robustness.
Table \ref{table:position} verifies that accelerometers situated on the forearm and biceps can be used with confidence to decode 5 upper-limb positions in the sagittal plane.
Despite encoding similar information from the ACC modality, the MED feature set consistently encoded positional information significantly better ($p<$0.05) than RMS.
Table \ref{table:gesture} provides an upper-limit of accuracy that can be achieved when position recognition is performed without fault.
Use of a sequential classification framework achieved no statistical difference between the within-position gesture recognition framework when using ACC MED to segment position and EMG TD to recognize gestures.
Although the position recognition performance of MED was statistically better than RMS, no statistical improvement is apparent in the gesture recognition accuracy of the sequential framework using these feature sets to decode position.

This study additionally corroborates the past outcomes of biomedical and HCI studies, where EMG is best for biomedical applications and ACC is best for HCI gesture recognition.
Gesture recognition for biomedical applications, such as prosthesis control, relies on class-separability provided through EMG features (96.3\%).
Although ACC features provide moderate class-separability for the WS (83.9\%) and WP classes (87.2\%), they provide only marginal class-separability for other classes (mean: 55.7\%).
HCI gesture recognition results found that ACC features substantially outperformed EMG features with the same set of gestures (HCI-A), a set of finger gestures (HCI-B), and a set of grasping gestures (HCI-C).
In contrast to past HCI experiments where 40 gestures are used together, the use of EMG TDPSD for gesture recognition with a subset of wrist gestures provided satisfactory accuracy (91.1\%).

Although findings were consistent with past studies, there remains a disconnect between the use of ACC for the recognition of gestures between the biomedical and HCI frameworks.
When no positional variance was purposely included (biomedical section of Table \ref{table:gesture}), ACC provided no real gesture-specific information resulting in low accuracy.
The high gesture recognition accuracy achieved using the HCI datasets is most likely an outcome of stratifying gestures across different positions to strategically reduce to improve the separability of the gestures.
This use of positional variance can be seen in Fig. \ref{fig:signals}, where the HCI dataset shows distinct changes in ACC signals during contractions that are uncharacteristic of mechanomyography.
This leveraging of positional variance was inferred in \cite{jiang2017feasibility}, where gestures performed ``in the air" resulted in higher accuracy than gestures performed when in contact with a surface.

A limitation of this study is the use of static contractions alone in the biomedical dataset. Past studies have found that including ramp contractions can reduce the impact of contraction intensity variability by incorporating more dynamics \cite{scheme2013training}. It is possible that there may exist repeatable ACC patterns during the transient segment of such ramp contractions that could be leveraged as part of future multi-modal myoelectric control systems. 

Ultimately, the consequence of different aims between biomedical and HCI applications can result in confusion when interpreting the outcomes of studies from both fields, especially the when terminology used to describe the gestures does not indicate the aim of the study. In light of this identified deficiency, it is suggested that a full review of past studies be conducted so as to develop a clear taxonomy and set of terminology that could be adopted by both of these expanding fields.


\bibliographystyle{unsrt}  

\bibliography{references}

\end{document}